\documentclass[aps,amsfonts,pra,twocolumn,superscriptaddress,showpacs]{revtex4}
\usepackage{epsfig,amsmath,amssymb,bm,epsf,graphics,psfrag} 
\def\bpd{\phi}
\def\flp{Q} 
\def\dll{{\cal N}} 
\def\lls{\xi^{2}} 
\def\ills{\xi^{-2}} 
\def\Ncontact{{N_{\rm c}}} 
\def\clp{{p}} 
\def\cln{{\cal P}} 
\def\hln{\widehat{\cln}} 
\def\arc{{\sigma}} 
\def\Rgy{{\cal R}_{\rm g}} 
\def\bit{a} 

\def\smthird{{\scriptstyle\frac{\scriptstyle 1}{\scriptstyle 3}}} 
\def\smhalf{{\scriptstyle\frac{\scriptstyle 1}{\scriptstyle 2}}} 
\begin{document} 
\title{Cavity Approach to the Random Solid State} 
\author{Xiaoming Mao} 
\affiliation{Department of Physics, 
University of Illinois at Urbana-Champaign, 
1110 W. Green St., Urbana, IL 61801, USA} 

\author{Paul M.~Goldbart} 
\affiliation{Department of Physics, 
University of Illinois at Urbana-Champaign, 
1110 W. Green St., Urbana, IL 61801, USA} 

\author{Marc M\'ezard} 
\affiliation{CNRS et Universit\'e Paris Sud, LPTMS, B\^at.\ 100, 91405 Orsay, France} 

\author{Martin Weigt} 
\affiliation{Institute for Scientific Interchange, 
Viale Settimio Severo 65, I-10133 Torino, Italy} 
% [CHECK] 
\date{June 7, 2005} 
% \date{\today} 

\begin{abstract} 
The cavity approach is used to address the physical properties 
of random solids in equilibrium.  Particular attention is paid 
to the fraction of localized particles and the distribution of  
localization lengths characterizing their thermal motion.  
This approach is of relevance to a wide class of random solids, 
including rubbery media (formed via the vulcanization of polymer 
fluids) and chemical gels (formed by the random covalent bonding 
of fluids of atoms or small molecules).  The cavity approach confirms results that have been obtained previously via replica mean-field theory, doing so in a way that sheds new light on their physical origin. 
\end{abstract} 
%
%[CHECK] 
% Curious: 4 pages if use just 05.70.Jk as PACS number.
%\pacs{05.70.Jk}%
%\pacs{82.70.Gg, 61.43.-j, 05.70.Jk}%
%---------------------------------
% 82.70.Gg  % Gels and sols
% 61.43.-j  % Disordered solids
% 05.70.Jk  % Critical point phenomena 
% 61.41.+e  % Polymers, elastomers, plastics
% 61.43.Er  % Other amorphous solids
% 64.70.Pf  % Glass transitions
%---------------------------------
\maketitle 

\noindent 
{\it Introduction\/}---% 
Permanent random chemical bonds, when introduced in sufficient number between the (atomic, molecular or macromolecular) constituents of a fluid, cause a phase transition---the vulcanization transition---to a new equilibrium state of matter: the random solid state.  In this state, some fraction of the molecules are spontaneously localized, and thus undergo thermal fluctuations about mean positions, the collection of which positions are random 
(i.e.~exhibit no long-range regularity).  
As this is a problem of statistical mechanics 
(the constituents are undergoing thermal motion) 
in the presence of quenched randomness 
(the constraints imposed by the random bonding), 
it has been addressed via the replica technique.  Indeed, the 
example of cross-linked macromolecular matter served as early 
motivation for Edwards' development of the replica technique 
(see, e.g., Ref.~\cite{REF:GGinSTG}).

Our purpose in this Letter is to consider two central diagnostics of the random solid state: the fraction $\flp$ of localized particles and the statistical distribution $\dll(\lls)$ of squared localization lengths $\lls$ of these localized particles.  Specifically, we show how results for these quantities can be obtained in an elementary and physically transparent way, via the cavity method.  The cavity method has proven flexible and powerful in the analysis of a variety of other disordered systems, e.g., spin glasses~[\onlinecite{REF:MPVbook}].  The present work is based on the version used to address spin glasses having finite connectivity~[\onlinecite{REF:MP2001-finite-conn}].

The results that we shall obtain via the cavity method are amongst those already known via a (less elementary and less physically transparent) application of the replica technique, together with a mean-field approximation; for reviews see Refs.~[\onlinecite{CGZ-AdvPhy},\onlinecite{G-Trieste}].  Thus, it was already known~[\onlinecite{castillo94}] that (as is typical for mean-field theories) $\flp$ obeys a transcendental equation, in this case 
\begin{equation} 
1-\flp=\exp (-\mu^{2}\flp ), 
\label{EQ:transcend}
\end{equation} 
where $\mu ^2$ is a parameter that controls the density of cross-links.  The instability of the fluid state and its replacement by the random solid state are signaled by the emergence, as $\mu$ is increased beyond a critical value (here unity), of a positive solution to Eq.~(\ref{EQ:transcend}), although the formulation is not restricted to the critical regime.  This result for $\flp$ was, in essence, found by Erd\"os and R\'enyi, in their classic work on the statistical properties of random graphs~[\onlinecite{REF:ER}].  As for the distribution $\dll(\lls)$, at the mean-field level and for near-critical values of the cross-link density, the replica approach yields the scaling form  
\begin{equation} \label{EQ:scaling}
% \dll(\lls)=({2}/{\epsilon \xi^4})\,\pi({2}/{\epsilon\lls}), 
  \dll(\lls)=\frac{2}{\epsilon \xi^{4}}\,\,
  \pi\!\left(\frac{2}{\epsilon\lls}\right), 
\end{equation} 
where $\epsilon$ [$\propto(\mu^{2}-1)$] measures the distance from the critical point, and in which the scaling function $\pi$ obeys the nonlinear integro-differential equation 
\begin{equation} 
\frac{d}{d\theta} 
\left( 
\smhalf\theta^{2}
\pi(\theta) 
\right) 
=\pi(\theta)-\left(\pi\circ\pi\right)(\theta), 
\label{EQ:IDE}
\end{equation} 
where $\pi\circ\pi$ indicates a Laplace convolution~\cite{REF:Normal}.
This result was obtained from a semi-microscopic approach by Castillo et al.~[\onlinecite{castillo94}] and re-derived via a
Landau-type theory by Peng et al.~[\onlinecite{peng98}].  
Some support for the results for $\flp$ and $\pi$ has been obtained numerically~[\onlinecite{barsky96}] and experimentally~[\onlinecite{REF:Dinsmore}].  

More recently, the mean-field level results have been improved in two directions: via renormalization-group analysis in the vicinity of the upper critical dimension (which is six for this random solidification transition)~[\onlinecite{peng00,REF:Janssen+Stenull}], and via analysis of the Goldstone fluctuations~[\onlinecite{REF:elast-swagatam}] (notably in two dimensions, where their consequences are---not surprisingly---dramatic).  We pause to mention that we view the cavity method as complementing rather than supplanting the replica method: the latter provides access to the powerful array of field-theoretic tools, the former opens the way to a perturbative treatment of correlations beyond mean-field. 

\begin{figure} 
\centerline{\psfig{figure=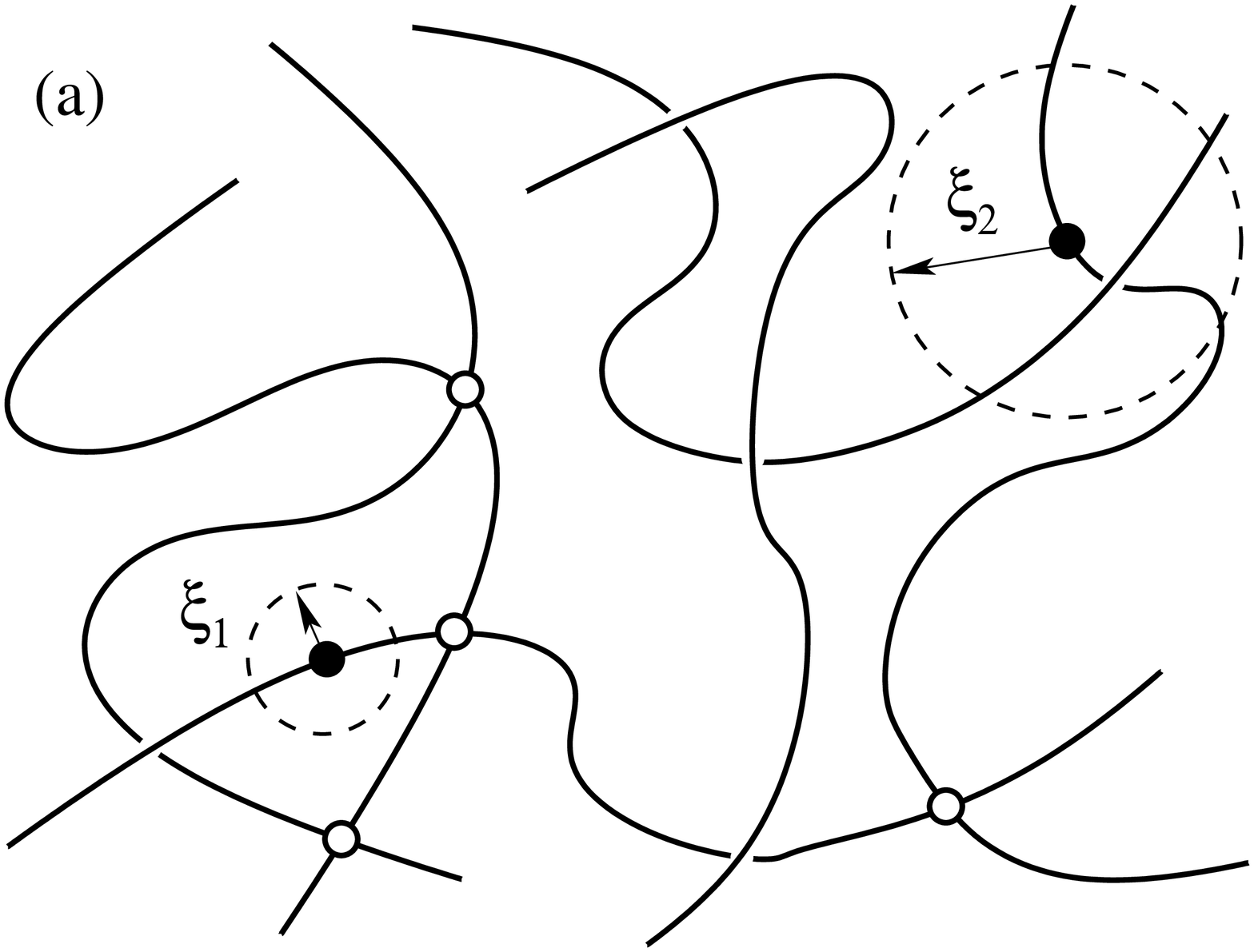,width=6cm,angle=0}}
\vspace{0.4cm} 
\centerline{\psfig{figure=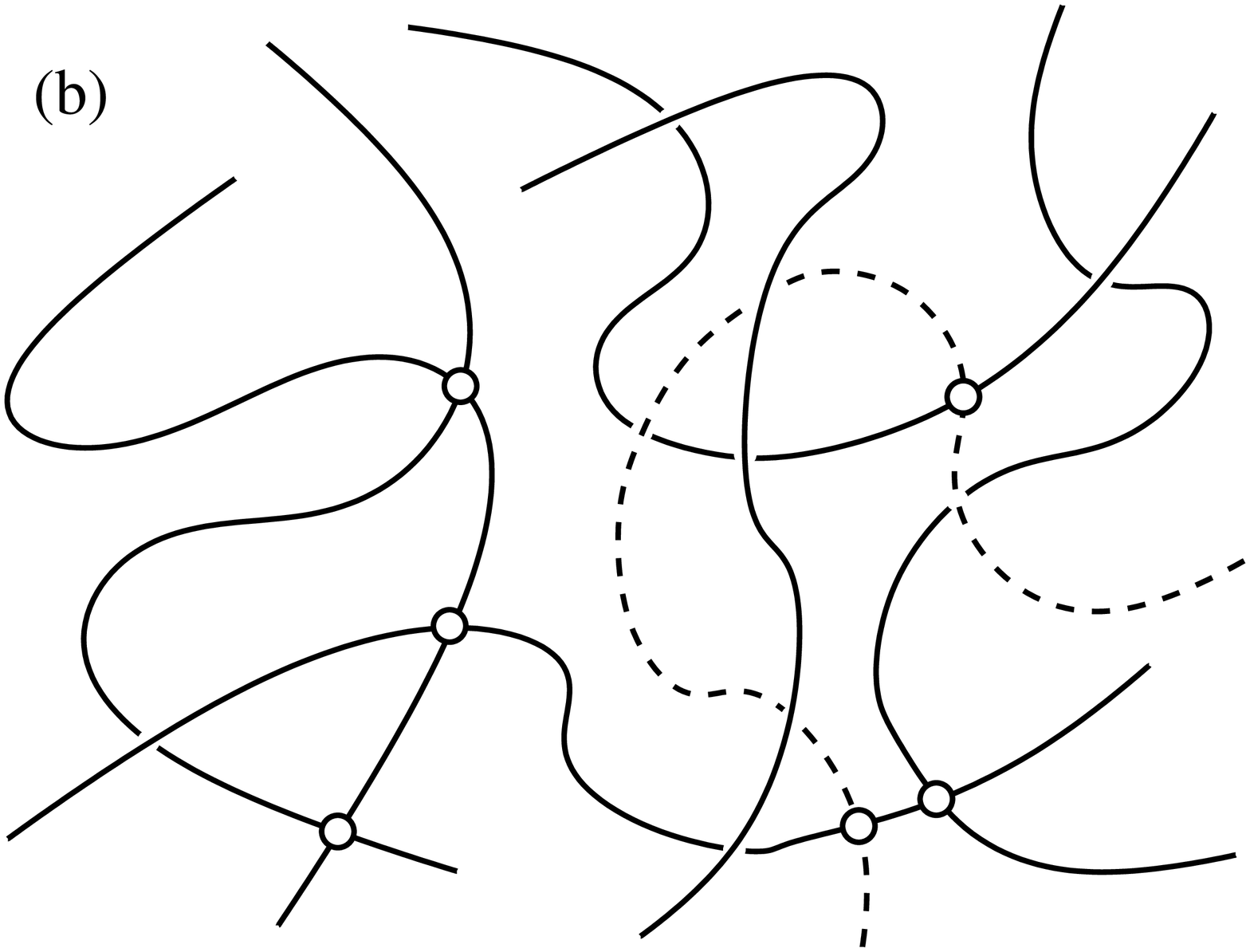,width=6cm,angle=0}}  
\caption{%
% (Color online) 
(a)~Snapshot of a cross-linked (open circles) macromolecular system, 
in which a fraction $Q$ of the macromolecules belong to the infinite 
cluster (and are thus localized), and the remaining fraction $1-Q$  
are delocalized.  A localized segment exhibits thermal position-fluctuations 
over a length-scale $\xi$ (i.e.~its localization length).  Due to the 
heterogeneity of the network, distinct segments have distinct localization 
lengths.  
(b)~Coupling of a new macromolecule (dashed) to the system in order to 
probe the distribution $\dll(\lls)$.  The new macromolecule has $\Ncontact$ 
contact points with the existing network, and cross-links are introduced 
independently at these points.} 
\label{fig:network}
\end{figure} 

\smallskip 
\noindent 
{\it Cavity method for randomly cross-linked macromolecules, 
especially in the vicinity of the random solidification transition\/}---% 
We begin by considering a system of vulcanized macromolecules, as depicted in Fig.~\ref{fig:network}a.  We characterize the system by the fraction $\flp$ of localized chain segments and the statistical distribution $\dll$ of mean squared localization lengths $\lls$ of the localized particles.  We then envisage adding a further macromolecule to the system, as shown in Fig.~\ref{fig:network}b.  Of all the segments on this chain, we suppose that a certain number $\Ncontact$ are sufficiently close to segments of the original system to have a chance of becoming cross-linked to them.  We suppose that fluctuations in this number are sufficiently small that we may neglect them.  
% [** CHECK physics of this condition?].   
Next, we consider a random cross-linking process that results, 
with independent probability $\clp$, in cross-links actually 
being introduced between each of the $\Ncontact$ close pairs.  
Within this framework, the probability that exactly $k$ 
cross-links are introduced is then given by the binomial formula: 
$\binom{\Ncontact}{k}
\clp^{k}(1-\clp)^{\Ncontact-k}$. 
We now ask the question: What is the probability $\cln_{k}$ that 
exactly $k$ of these cross-links are made to localized segments?  
To answer it, we make the approximation that the probabilities 
for the segments of the original system to be localized describe 
independent random variables, in which case the probability of 
any one such segment being localized is $\flp$.  
Then, collecting together the contributions to this probability, 
which arise from $k^{\prime}$ (with $k\le k^{\prime}\le\Ncontact$) 
cross-links being formed, of which exactly $k$ are to localized 
segments, we arrive at the formula 
\begin{equation} 
\cln_{k}=\sum_{k^{\prime}=k}^{\Ncontact} 
\binom{\Ncontact}{k^{\prime}}
\clp^{k^{\prime}}(1-\clp)^{\Ncontact-k^{\prime}} 
\binom{k^{\prime}}{k}
% \flp^{k^{\prime}}(1-\flp)^{\Ncontact-k^{\prime}}. 
  \flp^{k}(1-\flp)^{k^{\prime}-k}. 
\end{equation} 
Via a straightforward application of the binomial theorem one can perform this summation, and hence arrive at the result 
\begin{equation}\label{eq:p_k}
\cln_{k}= 
\binom{\Ncontact}{k}
(\clp\flp)^{k} 
(1-\clp\flp)^{\Ncontact-k}. 
\end{equation} 
Let us evaluate these probabilities for the three cases of relevance, viz., 
\begin{subequations} 
\begin{eqnarray} 
\cln_{0}&=&
(1-\clp\,\flp)^{\Ncontact}, 
\\ 
\cln_{1}&=&
\Ncontact\,\clp\,\flp(1-\clp\,\flp)^{\Ncontact-1}, 
\\ 
% \cln_{2}&=&\smhalf\Ncontact(\Ncontact-1)
% \clp\,\flp(1-\clp\,\flp)^{\Ncontact-2}. 
\cln_{2}&=&\smhalf\Ncontact(\Ncontact-1)
(\clp\,\flp)^{2}(1-\clp\,\flp)^{\Ncontact-2}. 
\end{eqnarray} 
\end{subequations} 
In the limit of main interest, viz.~$\clp\,\flp\ll 1$, 
these probabilities simplify to 
\begin{equation} 
\left(\cln_{0},\cln_{1},\cln_{2}\right)\approx
{\rm e}^{-\Ncontact\clp\flp}
\Big(1,\Ncontact\clp\flp,\smhalf \Ncontact(\Ncontact-1)(\clp\flp)^{2}\Big).
\end{equation} 

To arrive at a self-consistent equation for $\flp$ 
(as a function of $\clp$ and $\Ncontact$) 
we require that the probability of the added macromolecule being 
cross-linked to exactly zero localized segments be $1-\flp$, which gives 
\begin{equation} 
1-\flp=\left(1-\clp\,\flp\right)^{\Ncontact}. 
\end{equation} 
In the limit $\clp\,\flp\ll 1$ this becomes 
\begin{equation}\label{eq:Q_sc}
1-\flp=\exp\left(-\Ncontact\,\clp\,\flp\right), 
\end{equation} 
i.e., Eq.~(\ref{EQ:transcend}), provided we make the (physically sensible) identification $\clp\,\Ncontact\equiv\mu^{2}$. 

\begin{figure} 
\centerline{\psfig{figure=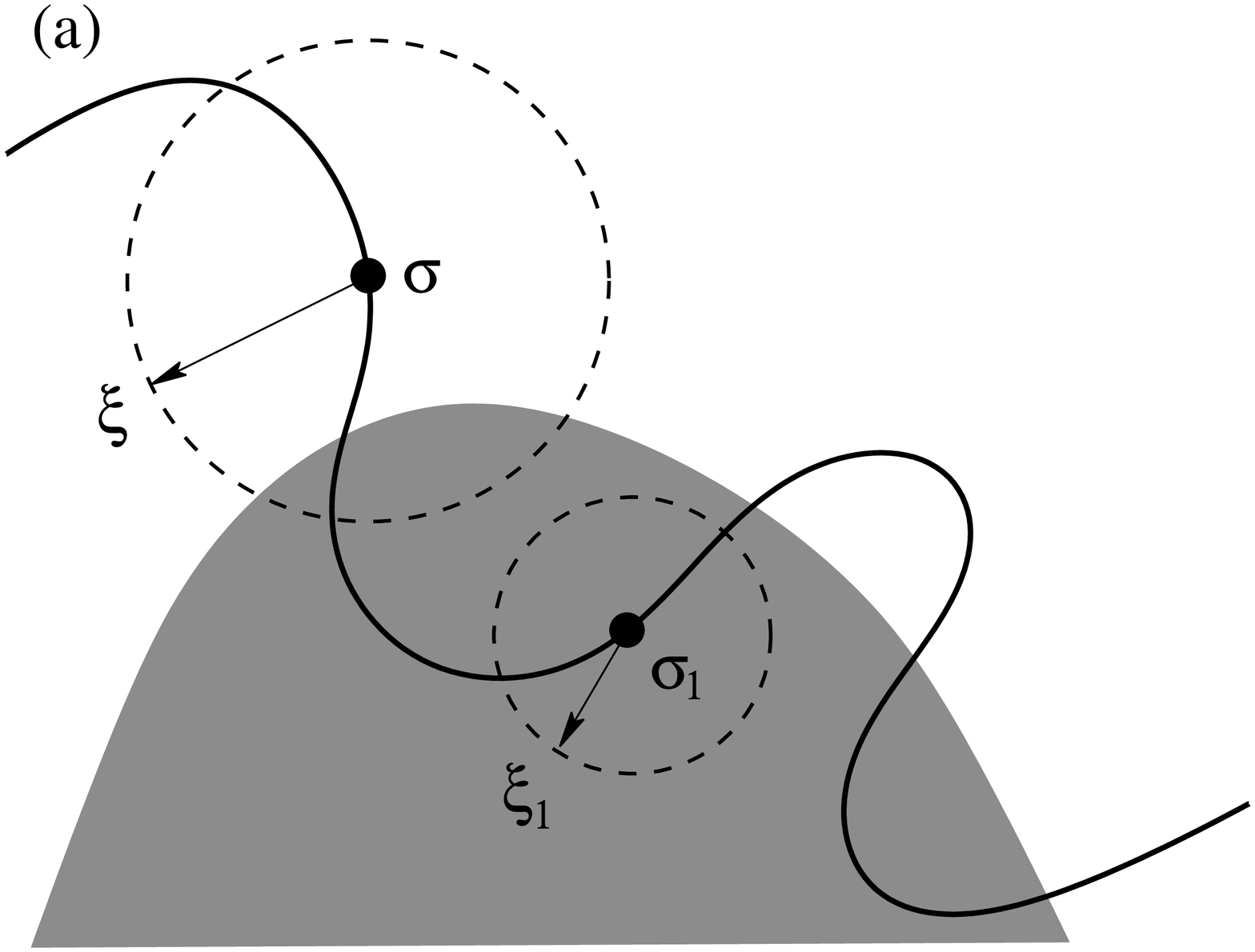,width=6cm,angle=0}}
\vspace{0.4cm} 
\centerline{\psfig{figure=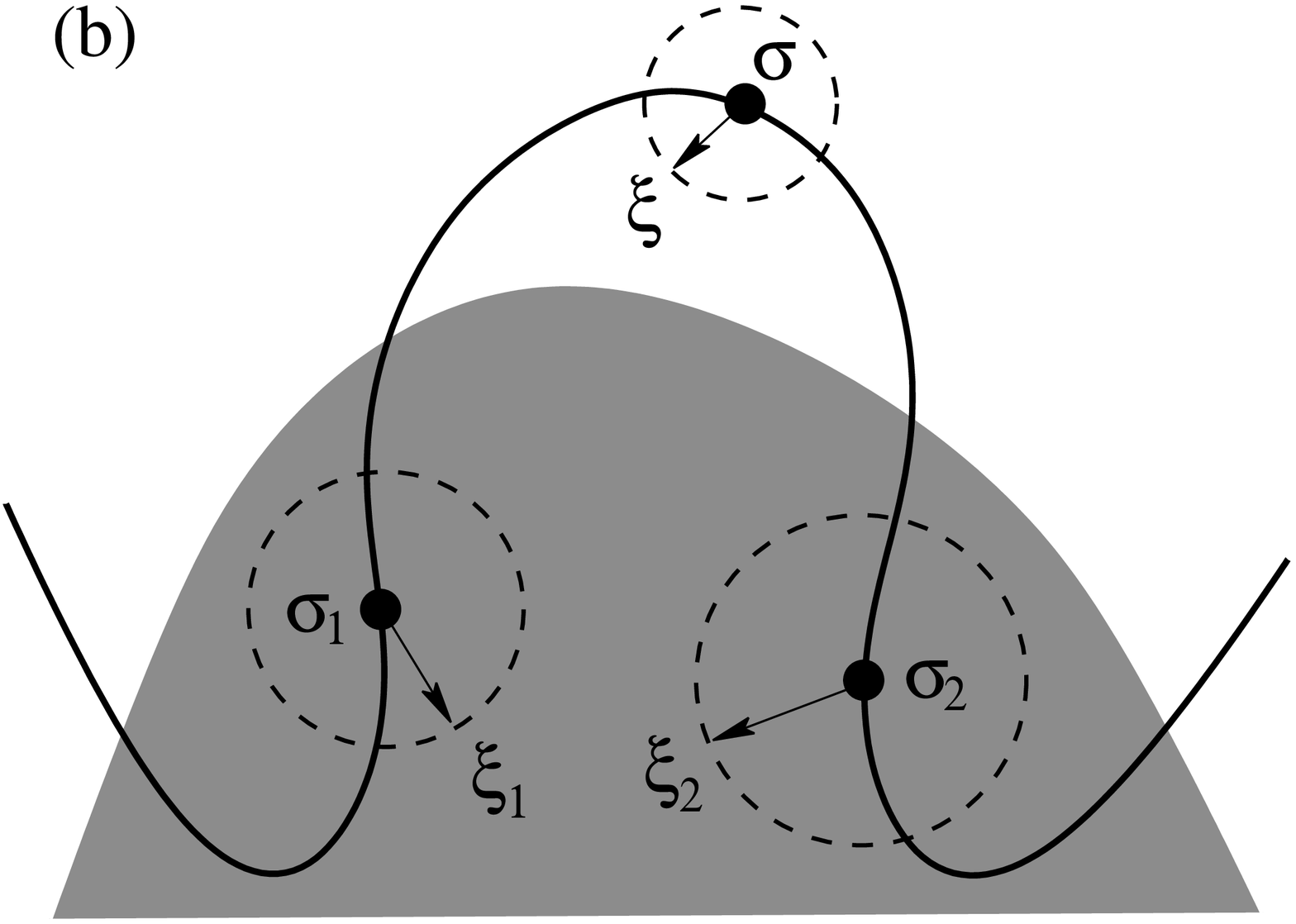,width=6cm,angle=0}}  
\caption{%
% (Color online) 
(a)~The added macromolecule is attached at a single segment to 
the infinite cluster (i.e.~the shaded region). 
(b)~The added macromolecule is attached at two segments to the 
infinite cluster.} 
\label{fig:addnew}
\end{figure} 

To arrive at a self-consistent equation for $\dll$ (as a function of $\clp$, $\Ncontact$ and $\lls$) we address the probability that the segment at arc-length $\arc$ (with $0\le\arc\le 1$) on the added macromolecule has squared localization length $\lls$.  As the segments of the added macromolecule will only be localized if they are attached to at least one localized segment of the original system, we should replace the probabilities $\{\cln_{k}\}_{k=0}^{2}$ by the probabilities conditioned on at least one attachment being to a localized segment of the original system (i.e.~$k=1,2$).  Thus we arrive at the probabilities $(\hln_{1},\hln_{2})$ which we write in the form $(1-\bit,\bit)$. 
Consider the situation in which the added macromolecule is attached at its arc-length $\arc_{1}$ to a {\it single\/} segment of the original system, that segment having squared localization length $\lls_{1}$, as depicted in Fig.~\ref{fig:addnew}a.  Furthermore, suppose that the added chain is Gaussian.  Then, by the elementary properties of random walks, the mean squared spatial separation of segments separated by arc-length $\vert{\arc-\arc_{1}}\vert$ is given by $\vert{\arc-\arc_{1}}\vert\,\Rgy^{2}$, 
where $\Rgy^{2}$ is the mean squared end-to-end distance of each chain.  Thus, if $\dll$ is the distribution for the squared localization length of the segment of the original system to which the new chain is attached at arc-length $\arc_1$, the distribution of the squared localization lengths $\lls$ for the segment at arc-length $\arc$ on the new chain will be given by 
\begin{equation} 
\int d\lls_{1}\,\dll(\lls_{1})\, 
\delta
\Big(\lls-
\left(\lls_{1}+\vert{\arc-\arc_{1}}\vert\,\Rgy^{2}\right)
\Big). 
\end{equation} 
Now, supposing that the addition to the squared localization length, 
$\vert{\arc-\arc_{1}}\vert\,\Rgy^{2}$, 
is small compared with the localization lengths that feature with appreciable weight in $\dll$, this approximates to 
\begin{equation} 
\dll\left(\lls-\vert{\arc-\arc_{1}}\vert\,\Rgy^{2}\right) 
\approx 
\dll(\lls) 
-\vert{\arc-\arc_{1}}\vert\,\Rgy^{2}\,\dll^{\prime}(\lls). 
\end{equation} 

Next consider the situation in which the added macromolecule is 
attached at its arc-lengths $\arc_{1}$ and $\arc_{2}$ to {\it two\/} segments of the original system, these segments having respective squared localization length $\lls_{1}$ and $\lls_{2}$, as shown in Fig.~\ref{fig:addnew}b. In fact, as the probability that the added chain has two cross-links to localized segments of the original system is (in the limit of interest) much smaller than the probability that it has one cross-link, we shall not need to keep track of the arc-length locations of the cross-links; in this situation it will be adequate to treat the added chain as a point object.  Then, as this object is attached to {\it two\/} localized objects, it is pinned more sharply than either, this {\it parallel\/} form of pinning giving rise to a {\it smaller\/} squared localization length $\lls$, via the formula  
\begin{equation} 
\ills=\ills_{1}+\ills_{2}. 
\end{equation} 
So, assuming that $\lls_{1}$ and $\lls_{2}$ are {\it independent\/} [and thus governed by the  joint distribution 
$\dll(\lls_{1})\,\dll(\lls_{2})$], the distribution of $\lls$ is given by 
\begin{equation} 
\int 
d\lls_{1}\,\dll(\lls_{1})\,
d\lls_{2}\,\dll(\lls_{2})\, 
\delta\left( 
\lls-\left(\ills_{1}+\ills_{2}\right)^{-1} 
\right). 
\end{equation} 

We now put these results together to construct the distribution of squared localization lengths for segment $\arc$ of the added 
chain, arriving at  
\begin{eqnarray} 
&& 
(1-\bit)
\Big(\dll(\lls) 
-\vert{\arc-\arc_{1}}\vert\,\Rgy^{2}\,\dll^{\prime}(\lls)
\Big) 
\nonumber\\ 
+\bit\!\!\!\!&&\!\!\!\!\int\!\!
d\lls_{1}\,\dll(\lls_{1})\,
d\lls_{2}\,\dll(\lls_{2})\, 
\delta\!
\left( 
\lls-\left(\ills_{1}+\ills_{2}\right)^{-1} 
\right)
\end{eqnarray} 
Finally, we average over the segment $\arc$ of the added chain, as well as the location $\arc_{1}$ of the cross-link, using $\int d\arc\,d\arc^{\prime}\,
\vert{\arc-\arc^{\prime}}\vert=1/3$, thus arriving at the self-consistent equation obeyed by the distribution of squared localization lengths: 
\begin{eqnarray} \label{EQ:unscaledDLL}
&& 
\dll(\lls)= 
(1-\bit)\left(\dll(\lls)-
\smthird
\,\Rgy^{2}\,\dll^{\prime}(\lls)\right) 
\nonumber\\ 
+\bit\!\!\!\!\!&&\!\!\!\!\int\! 
d\lls_{1}\,\dll(\lls_{1})\,d\lls_{2}\,\dll(\lls_{2})\, 
\delta\!\left(\lls-\left(\ills_{1}+\ills_{2}\right)^{-1}\right)
\label{eq:scdist} 
\end{eqnarray} 
Observe that by integrating both sides over $\lls$ and invoking the property that $\dll$ vanishes at the limits $\lls=0$ and $\infty$, we recover the normalization condition that $\int d\lls\,\dll(\lls)=1$.  The scaling property of $\dll$ shows up via the following change of dependent and independent variables:  
\begin{subequations} 
\begin{eqnarray} 
\lls&\to&\theta\equiv 
\frac{2}{3}\frac{1-\bit}{\bit}\,
\frac{\Rgy^{2}}{\lls}, 
\\ 
\dll(\lls)&\to&\pi(\theta)\equiv 
\frac{3}{2}
\frac{\bit}{1-\bit}
\,\frac{\lls}{\Rgy^{2}}\,\lls\,\dll(\lls),  
\end{eqnarray} 
\end{subequations} 
under which Eq.~(\ref{EQ:unscaledDLL}) becomes the 
sought integro-differential equation~(\ref{EQ:IDE}).  
We identify the parameter $\epsilon$ in Eq.~(\ref{EQ:scaling}) with $\frac{3\bit}{(1-\bit)\Rgy}$.  Notice that the cavity method allows us to compute corrections to Eq.~(\ref{EQ:IDE}) perturbatively in $1/\Rgy$.

\smallskip
\noindent 
{\it Cavity method for randomly bonded Brownian particles 
at arbitrary bonding densities\/}---% 
The cavity approach can be extended to address the chemical gelation transition, i.e., the transition triggered by the introduction of random covalent bonds between atoms or small molecules (rather than macromolecules) in the liquid state.  We shall address the model studied previously by Broderix et al.~\cite{BWZ-gelation}, which consists of a collection of point particles undergoing Brownian motion at a certain temperature.  Permanent bonds are then introduced at random between nearby particles, so that pairs of bonded particles become constrained softly (i.e.~by a spring-like, harmonic potential), so that the probability distribution $\bpd$ of their separations $\mathbf{r}_{j}-\mathbf{r}_{k}$ is Gaussian and characterized by a length-scale $l$: 
\begin{equation}
\bpd(\mathbf{r}_{j}-\mathbf{r}_{k})\propto\exp
\left(-
\vert\mathbf{r}_{j}-
     \mathbf{r}_{k}
\vert^{2}/2l^{2} 
\right) ,
\end{equation}
where $\mathbf{r}_{j,k}$ are the position vectors of particles $j$ and $k$.  
This model is sufficiently simple that the analysis of it need not be 
restricted to the critical region. 

To approach the statistics of this system of randomly bonded Brownian particles using the cavity method, we consider the process of adding a new particle.  The combinatorics of the bonding follows the form taken for the system of cross-linked macromolecules; we simply need to convert the notion of {\it contact points\/} into a spherical region of a certain radius in which the likelihood of particles being  bonded to one another is concentrated.  This sphere is centered on the new particle and, on average, includes $\Ncontact$ of the existing particles. Then, bonds are randomly introduced, with probability $\clp$, between the new particle and some of the $\Ncontact$ existing particles that are nearby.  Thus, the foregoing combinatorics continues to apply, and we arrive at the formula for the probability of having exactly $k$ bonds with the infinite cluster given in Eq.~(\ref{eq:p_k}).  As a consequence, we obtain the foregoing result for the fraction of the infinite cluster $Q$, Eq.~(\ref{eq:Q_sc}).

The physics of the localization lengths is, in fact, simpler 
for bonded Brownian particles.  When the new particle is connected 
via a spring of length-scale $l$ to {\it one\/} localized particle, 
its localization length $\Xi_{1}$ is given by 
$\Xi_{1}^{-2}=(\lls_1+l^2)^{-1}$.  
When it is connected {\it in parallel\/} via identical springs to 
$k$ localized particles, its localization length $\Xi_{k}$ is given by 
\begin{equation}
\label{eq:lls_bp}
\frac{1}{{\Xi}_{k}^{2}}\equiv\sum _{j=1}^{k} \frac{1}{\lls_{j}+l^2}.
\end{equation}

To construct the distribution of the squared localization length of the new particle, we shall average over all possible numbers of bonds, weighted with their corresponding probabilities. These  probabilities follow from the probabilities $\cln_{k}$ given by Eq.~(\ref{eq:p_k}), but normalized by a factor $Q^{-1}$ because the new particle will only be localized if it is bonded to at least one particle in the infinite cluster. Hence, we arrive at a self-consistency equation for localization-length distribution for the randomly bonded Brownian particle model: 
\begin{equation}
\label{eq:sc_lls_bp}
\dll (\lls)\!=\!
\sum_{k=1}^{\infty}\!
{\widehat\cln}_{k}\!\!
\int_{0}^{\infty}
\!\!\!\!\!
d\lls_1\dll(\lls _1)
\cdots
d\lls_k\dll(\lls _k)
\delta(\lls-\Xi_{k}^{2}),
\end{equation}
where the conditional probabilities ${\widehat\cln}_{k}$ 
are given by ${\widehat\cln}_{k}=\cln_{k}/\flp$ 
(for $k=1,2,3,\ldots$).
%%%
% \begin{equation}
% \label{p_k_c}{\widehat\cln}_{k}=\frac{1}{Q}
% \binom{\Ncontact}{k}(\clp\flp)^{k}\,
% (1-\clp\flp)^{\Ncontact-k}\quad(k=1,2,\ldots).
% \end{equation}
%%%
The distribution of localization lengths for the randomly bonded Brownian particle model was studied previously by Broderix et al.~\cite{BWZ-gelation} using the replica method and a Mayer cluster expansion.  To see that the cavity approach result~(\ref{eq:sc_lls_bp}) recovers their result, we take 
the limit $\Ncontact\to\infty$ whilst keeping finite the mean number of bonds from the new particle (either to the infinite cluster or to delocalized particles), i.e.~$\clp\Ncontact$.  In this limit, the binomial distribution 
% ${\widehat\cln}_{k}$ 
tends to a Poisson distribution:   
${\widehat\cln}_{k}
\longrightarrow
\left({(\clp\Ncontact)^{k}\,Q^{k-1}}/{k!}\right)
\textrm{e}^{-\clp\Ncontact Q}.$
The result of Broderix et al.~\cite{BWZ-gelation} then follows from Eq.~(\ref{eq:sc_lls_bp}) by 
(i)~transforming to a distribution for $\tau\equiv 1/\lls$, and 
(ii)~making the identifications $\kappa=l^{-2}$ and $c=\clp\Ncontact$ (i.e.~the mean of number of bonds associated with a single particle).

\smallskip 
\noindent 
{\it Acknowledgments\/}---% 
We thank H.~Castillo and A.~Zippelius for valuable discussions. PG, MM and MW thank for its hospitality the Kavli Institute for Theoretical Physics at the University of California--Santa Barbara, where this work was begun. PG also thanks for its hospitality the Laboratoire de Physique Th\'eorique et Mod\`eles Statistiques at 
l'Universit\'e Paris Sud, where it was completed. 
This work was supported by  
NSF DMR02-05858 (XM, PMG)
NSF  PH99-07949 (PG, MM, MW).


\begin{thebibliography}{99}
\bibitem{REF:Normal}
As $\pi$ is a distribution it must be non-negative and normalized: $\int_0 ^\infty d\theta\,\pi(\theta)=1$.  Equation~(\ref{EQ:IDE}) is, however, self-normalizing, in the sense that solutions that behave reasonably, as $\theta$ tends to zero or infinity, are automatically normalized, as integrating the equation between these limits, term by term, reveals.
\end{thebibliography}
\end{document}